\documentclass[journal=jpclcd,manuscript=letter]{achemso}

\usepackage[version=3]{mhchem} 
\usepackage[inline]{enumitem}
\usepackage{subcaption}        
\usepackage{xcolor}
\usepackage{booktabs}
\usepackage{multirow}
\usepackage{hyperref}

\author{Jingjing Li}
\affiliation{Key Laboratory of Theoretical and Computational Photochemistry, Ministry of Education, College of Chemistry, Beijing Normal University, Beijing 100875, People’s Republic of China}
\author{Weitang Li}
\affiliation{School of Science and Engineering, The Chinese University of Hong Kong, Shenzhen, 518172, P. R. China.}
\author{Xiaoxiao Xiao}
\author{Limin Liu}
\author{Zhendong Li}
\author{Jiajun Ren}
\email{jjren@bnu.edu.cn}
\author{Weihai Fang}
\affiliation{Key Laboratory of Theoretical and Computational Photochemistry, Ministry of Education, College of Chemistry, Beijing Normal University, Beijing 100875, People’s Republic of China}

\title{Multi-set variational quantum dynamics algorithm for simulating nonadiabatic dynamics on quantum computers}

\begin{document}

\begin{tocentry}
\begin{figure}[H]
    \centering
    \includegraphics[width=\textwidth]{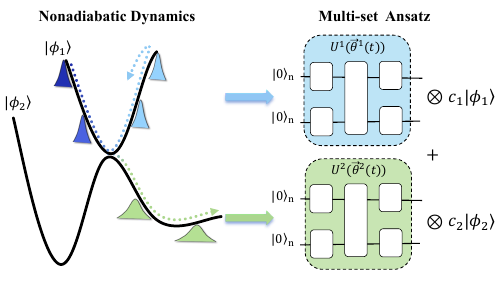}
\end{figure}

\end{tocentry}

\begin{abstract}
    Accelerating quantum dynamical simulations with quantum
    computing has received considerable attention but remains a
    significant challenge. In variational quantum algorithms for
    quantum dynamics, designing an expressive and shallow-depth
    parameterized quantum circuit (PQC) is a key difficulty.
    Here, we proposed a multi-set variational quantum dynamics
    algorithm (MS-VQD) tailored for nonadiabatic dynamics
    involving multiple electronic states. MS-VQD employs
    multiple PQCs to represent the electronic-nuclear coupled
    wavefunction, with each circuit adapting to the motion of nuclear
    wavepacket on a specific potential energy surface. By
    simulating excitation energy transfer dynamics in molecular
    aggregates described by the Frenkel-Holstein model, we
    demonstrated that MS-VQD achieves the same accuracy as
    traditional VQD while requiring significantly shallower
    PQCs. Notably, its advantage increases with the number of
    electronic states, making it suitable for simulating
    nonadiabatic quantum dynamics in complex molecular systems.
\end{abstract}


Nonadiabatic dynamics play a crucial role in chemistry, physics,
and material science, influencing processes such as
photochemical reactions, excitation energy transfer, charge
separation, and relaxation.~\cite{yarkony2012nonadiabatic,long2017nonadiabatic,
jang2018delocalized, dimitriev2022dynamics} In these processes,
quantum effects, including coherence and tunneling, have been
identified essential.~\cite{bredas2017photovoltaic,
wang2019quantum, schultz2024coherence} However, accurately
simulating nonadiabatic dynamics with quantum effects remains a
major challenge in theoretical
chemistry.~\cite{tully2012perspective, curchod2018ab} In
classical computation, exact methods for solving the coupled
electronic-nuclear time-dependent Schrödinger equation (TDSE)
suffer from exponential growth in complexity and memory with
increasing system size, known as the ``quantum exponential
wall''. Over the past few decades, various approximate but
numerically exact methods have been developed to overcome this
problem and have made great success.~\cite{makri1998quantum,
worth2008using, tanimura2020numerically, ren2022time} However,
their scalability and general applicability remain to be
carefully assessed.

The advent of quantum computing offers compelling new avenues to
overcome these challenges.~\cite{cao2019quantum,bauer2020quantum,
mcardle2020quantum, ollitrault2021molecular, motta2022emerging, 
delgado2024quantum,fauseweh2024quantum} Simulating TDSE is considered one of the
most promising applications for demonstrating quantum
advantage.~\cite{lloyd1996universal} To this end, various
quantum algorithms have been proposed,~\cite{miessen2023quantum}
including decomposition algorithms and variational
algorithms.~\cite{lloyd1996universal, Childs2012HamiltonianSU, berry2015simulating,
li2017efficient,yuan2019theory,campbell2019random,
cirstoiu2020variational,hu2020quantum,
barison2021efficient,schlimgen2022quantum,
wan2024hybrid} Their applications to different chemical dynamics
problems have also emerged, for both closed and open
systems.~\cite{kassal2008polynomial,macridin2018electron,
endo2020variational, ollitrault2020nonadiabatic,
lee2022simulating, tazhigulov2022simulating,
kovyrshin2023nonadiabatic, zhang2023quantum,gomes2023computing,
luo2024variational,li2024toward, lan2024integrating,
lyu2024simulating,gallina2024simulating, walters2025variational,
dan2025simulating} Compared to decomposition algorithms, hybrid
quantum-classical variational algorithms require significantly
shallower circuit depths and exhibit greater resilience to noise,
making them more practical for implementation on near-term
quantum hardware. Their feasibility has been validated in recent experimental demonstrations.~\cite{chen2020demonstration,
lee2022simulating, gomes2023computing} 

In variational quantum algorithms,~\cite{cerezo2021variational,tilly2022variational} the wavefunction ansatz, implemented as a parameterized quantum circuit (PQC), plays a crucial role in determining both accuracy and efficiency.
An effective PQC must satisfy two key criteria: (i)
high expressivity, enabling it to represent highly entangled
quantum states beyond classical ansatzes, and (ii) short circuit
depth, ensuring practical implementation on the current noisy quantum
hardware. To meet these requirements, various PQC designs have
been proposed,~\cite{peruzzo2014variational, wecker2015progress,
kandala2017hardware,lee2018generalized,grimsley2019adaptive,sim2019expressibility,
zhang2022variational,fan2023quantum, zeng2023quantum,
xiao2024physics,li2025quantum} including the unitary coupled
cluster ansatz for
electronic structure problems,~\cite{peruzzo2014variational} the
Hamiltonian variational ansatz,~\cite{wecker2015progress}
hardware-efficient ansatz,~\cite{kandala2017hardware} and
quantum-classical hybrid ansatz,~\cite{zhang2022variational,
li2025quantum} etc.

\begin{figure*}[htb]
    \centering
    \includegraphics[width=0.99\textwidth]{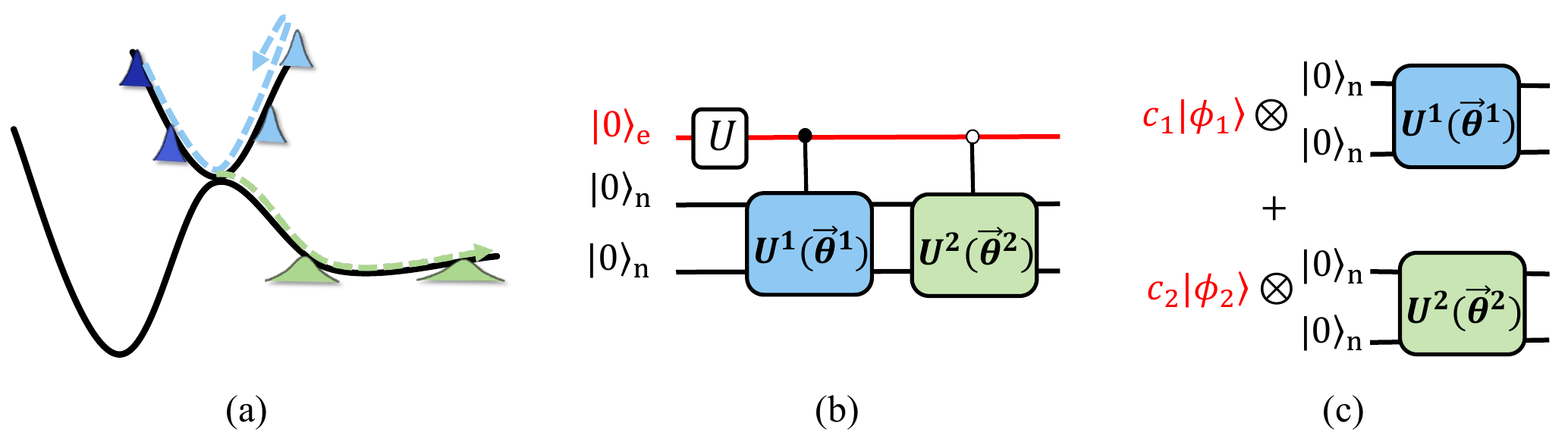}
    \caption{
        (a) Schematic illustration of a two-state nonadiabatic
        dynamics process. One wavepacket splits into two
        upon passing through a strong nonadiabatic coupling
        region.
        (b) In variational quantum dynamics, two distinct ansatz blocks
            are employed, determined by the electronic state index,
            resulting in a doubled circuit depth.
        (c) In multi-set variational quantum dynamics, two separate
            circuits are used, each representing the nuclear wavepacket
            on a specific potential energy surface.
    }
    \label{fig:pes}
\end{figure*}

Current variational quantum dynamics (VQD) algorithms use a
single PQC $U(\vec \theta(t))$ to represent the wavefunction of the entire system,~\cite{yuan2019theory,ollitrault2021molecular}
\begin{gather}
    |\Psi(t) \rangle_\textrm{VQD} = U(\vec \theta(t)) |\mathbf
    0 \rangle.
\end{gather}
In nonadiabatic dynamics involving multiple electronic
states, $U(\vec \theta(t))$ entangles both electronic and nuclear degrees of freedom. However, when potential energy
surfaces (PESs) have distinct landscapes, the motion of nuclear wavepacket
on each PES varies significantly, making it difficult for
a single compact PQC to accurately capture the full dynamics.
A schematic example is shown in Figure~\ref{fig:pes}(a), where
an initial nuclear wavepacket splits into two after passing
through a strong nonadiabatic coupling region. On the steeper
upper PES, the wavepacket reflects backward, while on the
flatter lower PES, it propagates forward. 
In order to accurately describe the two wavepackets,
the PQC should incorporate two distinct ansatz blocks $U^1(\vec \theta^1(t))$ and $U^2(\vec \theta^2(t))$, controlled by the qubit encoding the electronic state
(Figure~\ref{fig:pes}(b)), resulting in a doubled circuit depth.

To address this challenge, in this study, we proposed a multi-set variational
quantum dynamics algorithm (MS-VQD), an extension of the
original VQD framework tailored for multi-state dynamics,
inspired by the classical multi-set tensor network
ansatz.~\cite{fang1994multiconfiguration} 
We demonstrated the effectiveness of MS-VQD in simulating
excitation energy transfer (EET) dynamics in molecular
aggregates described by the Frenkel-Holstein model. EET plays a
fundamental role in light-harvesting complexes and organic solar
cells.~\cite{jang2018delocalized, bredas2017photovoltaic} Our results showed that MS-VQD achieves the same accuracy
as conventional VQD while requiring significantly shallower
PQCs. Notably, its advantage becomes more pronounced as the
number of electronic states increases, making it particularly
well-suited for complex nonadiabatic quantum dynamics
simulations.

The wavefunction ansatz of MS-VQD is
\begin{align}
    |\Psi(t) \rangle_\textrm{MS-VQD} & = \sum_p c_p(t)|\phi_p\rangle_\textrm{e}
    |\chi_p(t)\rangle_\textrm{n},
       \\
    | \chi_p(t) \rangle_\textrm{n} & = U^p(\vec \theta^p(t)) |\mathbf
    0\rangle_\textrm{n},
\end{align}
where the subscript e/n indicates electronic/nuclear wavefunction. MS-VQD employs separate PQCs, $U^p(\vec{\theta} ^p(t))$, each independently
representing the nuclear wavepacket $|\chi_p \rangle$
associated with the $p$th electronic state $|\phi_p\rangle$.
These wavepackets are then multiplied by their respective
electronic states and linearly combined to reconstruct the full
wavefunction (Figure~\ref{fig:pes}(c)). 
This structure allows each PQC to better adapt to its corresponding wavepacket,
resulting in a more compact and efficient ansatz.

The nonadiabatic Hamiltonian in the diabatic representation can
be generally expressed as 
\begin{gather}
    \hat H  = \hat T + 
    \begin{bmatrix}
        \hat V_{11}  &\hat V_{12} &\cdots &\hat V_{1N} \\ 
        \hat V_{21}  &\hat V_{22} &\cdots &\hat V_{2N} \\ 
        \cdots          & \cdots         &\cdots & \cdots \\
       \hat V_{N1}  &\hat V_{N2} &\cdots &\hat V_{NN} \\ 
    \end{bmatrix}.
\end{gather}
The diabatic energies and coupling terms depend on the
nuclear coordinates $\hat V_{pq} \equiv	 \hat V_{pq}(\vec R)$.

\begin{figure*}[htb]
    \centering
    \includegraphics[width=1\linewidth]{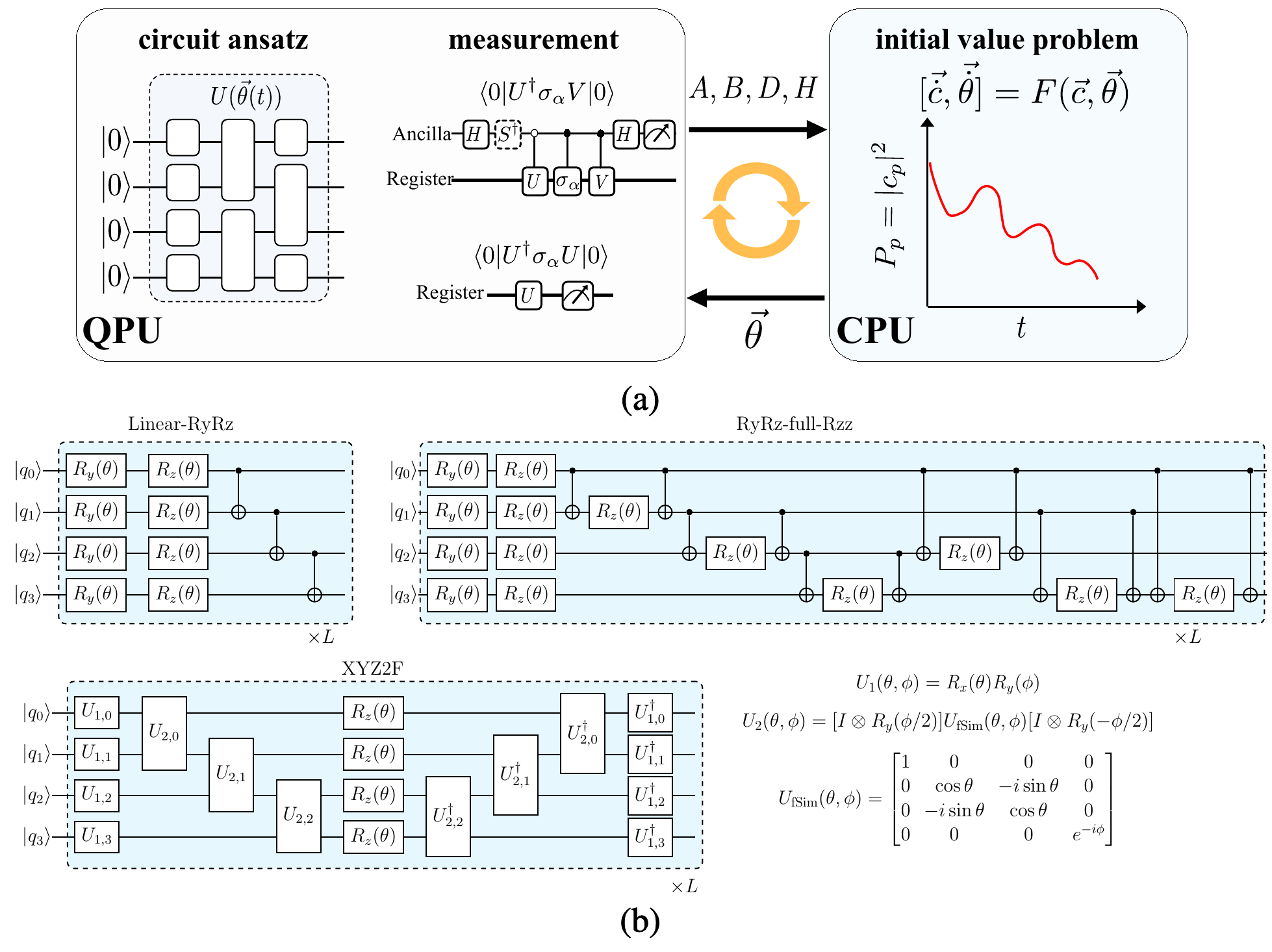}
    \caption{ 
    (a) Workflow of the hybrid quantum-classical multi-set
    variational quantum dynamics algorithm. The quantum computer
    prepares the nuclear wavepacket and measures the matrix
    elements using direct measurement or the Hadamard test
    algorithm. The classical computer updates the parameters by
    solving an initial value problem.
    (b) Three types of ansatzes used in this study: linear-RyRz
    ansatz, RyRz-full-Rzz ansatz, and XYZ2F ansatz.
    }
    \label{fig:workflow}
\end{figure*}

Based on the McLachlan’s time-dependent variational principle
$\delta || (\hat H-i\partial/\partial t)|\Psi(\vec{\theta}(t))
\rangle ||^2=0$,~\cite{mclachlan1964variational} the equations of motion of MS-VQD is
\begin{gather}
    \sum_k |c_p|^2 \mathrm{Re}(A_{lk}^{pp}-
    D_l^{pp}D_k^{pp^*})\dot{\theta}^p_k  = \mathrm{Im}\sum_q
    c_p^*c_q(B_l^{pq}-H_{pq}D_l^{pp})
     \label{eq:theta_deri}  \\
    \dot{c}_p+c_p\sum_k D_k^{pp^*}\dot{\theta}^p_k+i\sum_q c_q
    H_{pq} = 0, 
    \label{eq:c_deri}
\end{gather}
where the matrices $\mathbf A, \mathbf B, \mathbf D,
\mathbf H$ are:
\begin{align}
    A_{lk}^{pp} & = \langle\frac{\partial\chi_p}{\partial\theta^p_l}
            |\frac{\partial\chi_p}{\partial\theta^p_k}\rangle \\     
    B_l^{pq} & = \langle\frac{\partial\chi_p}{\partial\theta^p_l}
            |\hat T + \hat V_{pq} |\chi_q \rangle \\
    D_k^{pp} & = \langle\frac{\partial\chi_p}{\partial\theta^p_k}|\chi_p\rangle \\
    H_{pq} & = 
    \langle \chi_p | \hat T + \hat V_{pq} |\chi_q \rangle.
\end{align}
The detailed derivation is provided in the Appendix. The matrix
elements of $\mathbf{A}, \mathbf{B}, \mathbf{D}, \mathbf{H}$ can
be measured using either the direct measurement or the Hadamard
test algorithm, as illustrated in Figure~\ref{fig:workflow}(a). The
details are presented in Section~1 and Figure~S1 of Supporting Information (SI).  
When only a single electronic state is present, eqs~\eqref{eq:theta_deri} and \eqref{eq:c_deri} reduce to a simplified form,
\begin{gather} 
    \sum_k \mathrm{Re}(A_{lk}-D_lD_k^*)\dot{\theta}
    _k=\mathrm{Im}(B_l-ED_l), \label{eq:singleset}\\
   \dot{c} + c\sum_k D_k \dot{\theta}_k + i c E = 0.
\end{gather}
Eq~\eqref{eq:singleset} is consistent with the equation of motion for traditional VQD, as
derived in the previous studies.~\cite{yuan2019theory}
At first glance, MS-VQD appears more complex than VQD due to the
additional index $p$, which corresponds to a specific electronic
state. However, the left-hand side of the linear equation
eq~\eqref{eq:theta_deri} depends only on $p$, resulting in a
block-diagonal structure. This simplifies solving
the linear equation on a classical computer.
The overall processes of MS-VQD is illustrated in
Figure~\ref{fig:workflow}(a), including:
\begin{enumerate}[label=(\roman*)]
  \item Initialize $\vec c$ and $\vec \theta$ according to
      the initial condition;
  \item Prepare PQC on quantum computer and measure the
      matrix elements of $\mathbf A, \mathbf B, \mathbf D,
        \mathbf H$; \label{item:1}
  \item Calculate the time derivative $\vec{\dot \theta}$ and
      $\vec{\dot c}$ according to eqs~\eqref{eq:theta_deri}
        \eqref{eq:c_deri} and update $\vec \theta$ and $\vec c$
        with any proper solver for initial value problems on
        classical computer. Return back to \ref{item:1}.
\end{enumerate}

To evaluate the effectiveness of MS-VQD in simulating
nonadiabatic quantum dynamics, we applied it to model the
excitation energy transfer dynamics in one-dimensional molecular
aggregates using the Frenkel-Holstein model Hamiltonian, which is
\begin{align}
    \hat H = & \sum_p  \varepsilon_p
    |\phi_p\rangle\langle\phi_p| +\sum_{p} J
    (|\phi_p\rangle\langle\phi_{p+1}| +
    |\phi_{p+1}\rangle\langle\phi_{p}|)\nonumber \\
    & + \sum_{pm}
    \omega_m \hat b_{pm}^\dagger \hat b_{pm} + \sum_{pm} g_m
    \omega_m 
   |\phi_p\rangle\langle\phi_p| (\hat b_{pm}^\dagger + \hat
    b_{pm}).
\end{align}
In this model, $\varepsilon_p$ denotes the excitation energy of
the local electronic state $|\phi_p \rangle$, $J$ denotes the
excitonic coupling. $\hat{b}_{pm} ^\dagger$ and $\hat{b}_{pm}$
correspond to the creation and annihilation operators of the
$m$th vibrational mode of the $p$th molecule, with frequency
$\omega_m$ and dimensionless electron-vibration coupling
constant $g_m$. 

To encode the infinitely large bosonic Hilbert space into qubits,
we truncated the local bosonic energy levels to a finite size
$d$ and mapped them to qubit states using Gray code. For example,
when $d=4$, the encoding follows $|00\rangle \to |0\rangle$,
$|01\rangle \to |1\rangle$, $|11\rangle \to |2\rangle$, and
$|10\rangle \to |3\rangle$. Gray code has been shown to not only
reduce the number of qubits compared to unary encoding but also
reduce the number of two-qubit gates needed to represent
$b^\dagger$/$b$ compared to standard binary
encoding.~\cite{sawaya2020resource} 
In addition, MS-VQD is also compatible with the recently
proposed variational encoding method, which can further reduce
qubit requirements.~\cite{li2023efficient}
Besides encoding the bosonic state, in traditional VQD,
electronic states are encoded using binary code. 
For a system with $N$ molecules and $K$ nuclear degrees of
freedom per molecule, since MS-VQD does not explicitly encode
electronic states on the quantum computer,
the number of qubits is reduced from $NK \log_2 d +
\log_2 N$ in VQD to $NK \log_2 d$.
Meanwhile, the number of Hamiltonian terms after
mapping into qubits in MS-VQD is also much less than that in VQD shown in
Figure~S2. 

In the choice of ansatz for each PQC in both VQD and MS-VQD, any
proper ansatz can be used. We employed three types: the
linear-RyRz ansatz,~\cite{sim2019expressibility} the RyRz-full-Rzz ansatz, and the XYZ2F
ansatz,~\cite{xiao2024physics} as shown in Figure~\ref{fig:workflow}(b).
The first two are simple hardware efficient ansatzes, where
single-qubit rotation gates and two-qubit CNOT/Rzz gates are
placed in alternation. The third, XYZ2F, is a recently developed
ansatz~\cite{xiao2024physics} that satisfies key physical
constraints and has been proven to be universal, systematically
improvable, and size-consistent. In our calculations, the number
of layers $L$ of these ansatzes are gradually increased to
enhance the expressivity.

We primarily focused on the exciton population dynamics, defined as
$P_p(t) = |\langle \phi_p | \Psi(t) \rangle|^2$. The initial
condition assumes that the exciton is localized on the first
molecule, with all vibrational modes in their ground state. In
MS-VQD, $P_p = |c_p|^2$, which can be directly computed
classically, whereas in VQD, $P_p$ is obtained by sampling the
qubits encoding the electronic state, expressed as $P(j_1 j_2
\cdots, p=\sum_k j_k 2^{k})$. All simulations were performed
without noise using the quantum simulator TensorCircuit~\cite{zhang2023tensorcircuit}
and TenCirChem~\cite{li2023tencirchem}. 
Reference data was calculated
using the numerically exact time-dependent density matrix
renormalization group algorithm (TD-DMRG).~\cite{ren2022time}

\begin{figure}[htb]
    \centering
    \includegraphics[width=0.8\linewidth]{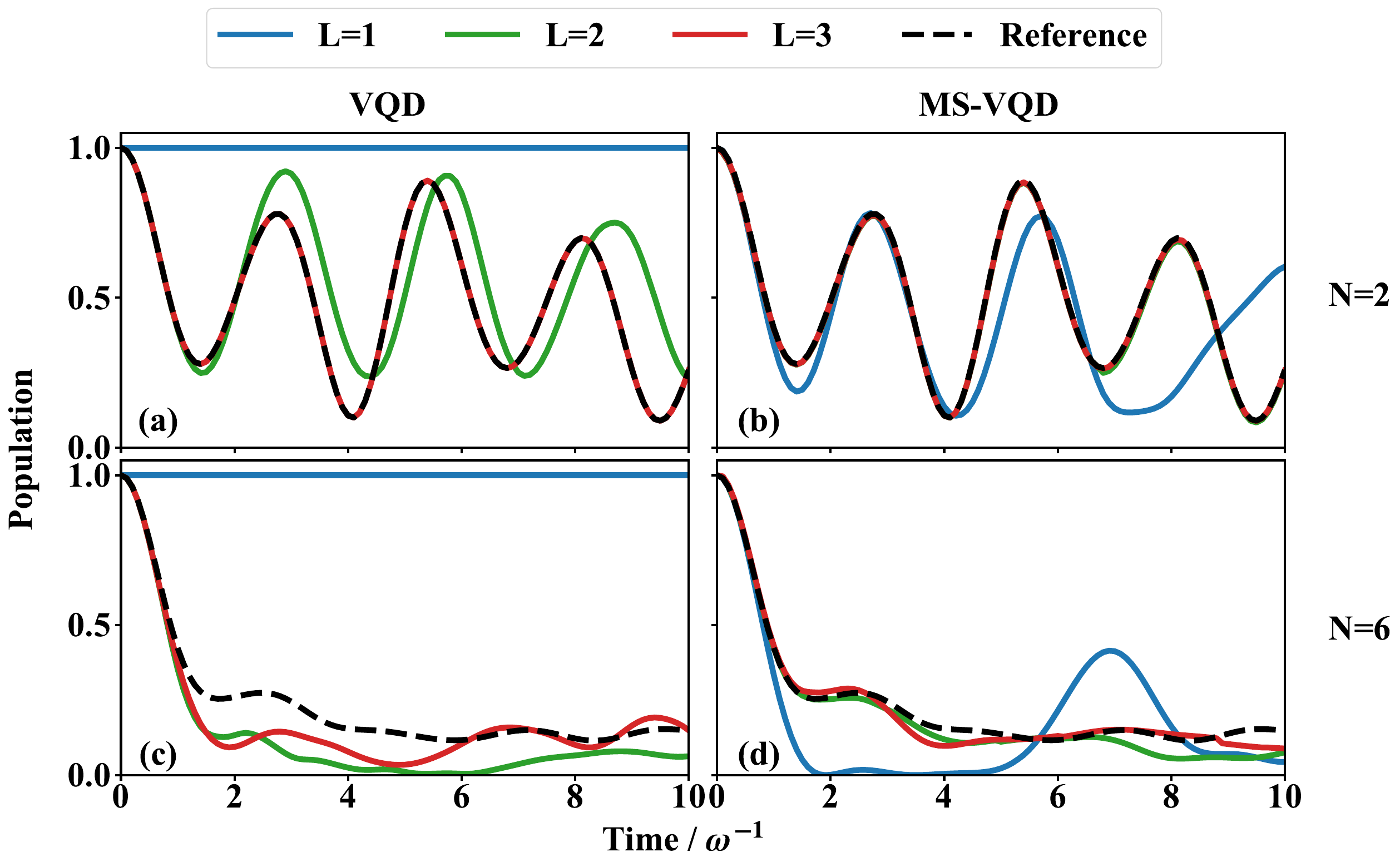}
    \caption{
        The exciton population on the first molecule as a function
        of time, $|\langle \phi_1 | \Psi(t)\rangle|^2$. (a) and (b)
        Results of VQD (a) and MS-VQD (b) for the dimer model.
        Different colors represent different layers of the ansatz,
        with $L = 1, 2, 3$. The dashed black curve shows the
        reference results calculated by the time-dependent density
        matrix renormalization group algorithm. (c) and (d) Similar to
        (a) and (b), but for the hexamer model. The linear-RyRz
        ansatz is used.
    }
    \label{fig:pop_vs_time}
\end{figure}

Figure~\ref{fig:pop_vs_time} presents the exciton population
$P_1$ of the first molecule for dimer ($N=2$) and hexamer
($N=6$), using the linear-RyRz ansatz with $L=1,2,3$
layers. 
One vibrational mode per molecule is considered.
The parameters used are $g=1$, $J=-1$, $d=2$,
$dt=0.1$, with $\omega=1$ as the unit. Comparing
Figure~\ref{fig:pop_vs_time}(a) to (c) and (b) to (d), we
observed that as the number of molecules increases, the
simulation becomes more challenging, requiring additional ansatz
layers to maintain accuracy for both VQD and MS-VQD.
Compared to VQD, MS-VQD consistently achieves
higher accuracy with the same $L$, regardless of system size.
Similar behaviors were observed for the other two ansatzes, as
shown in Figure~S3 and S4.

\begin{figure}[htb]
    \centering
    \includegraphics[width=0.95\linewidth]{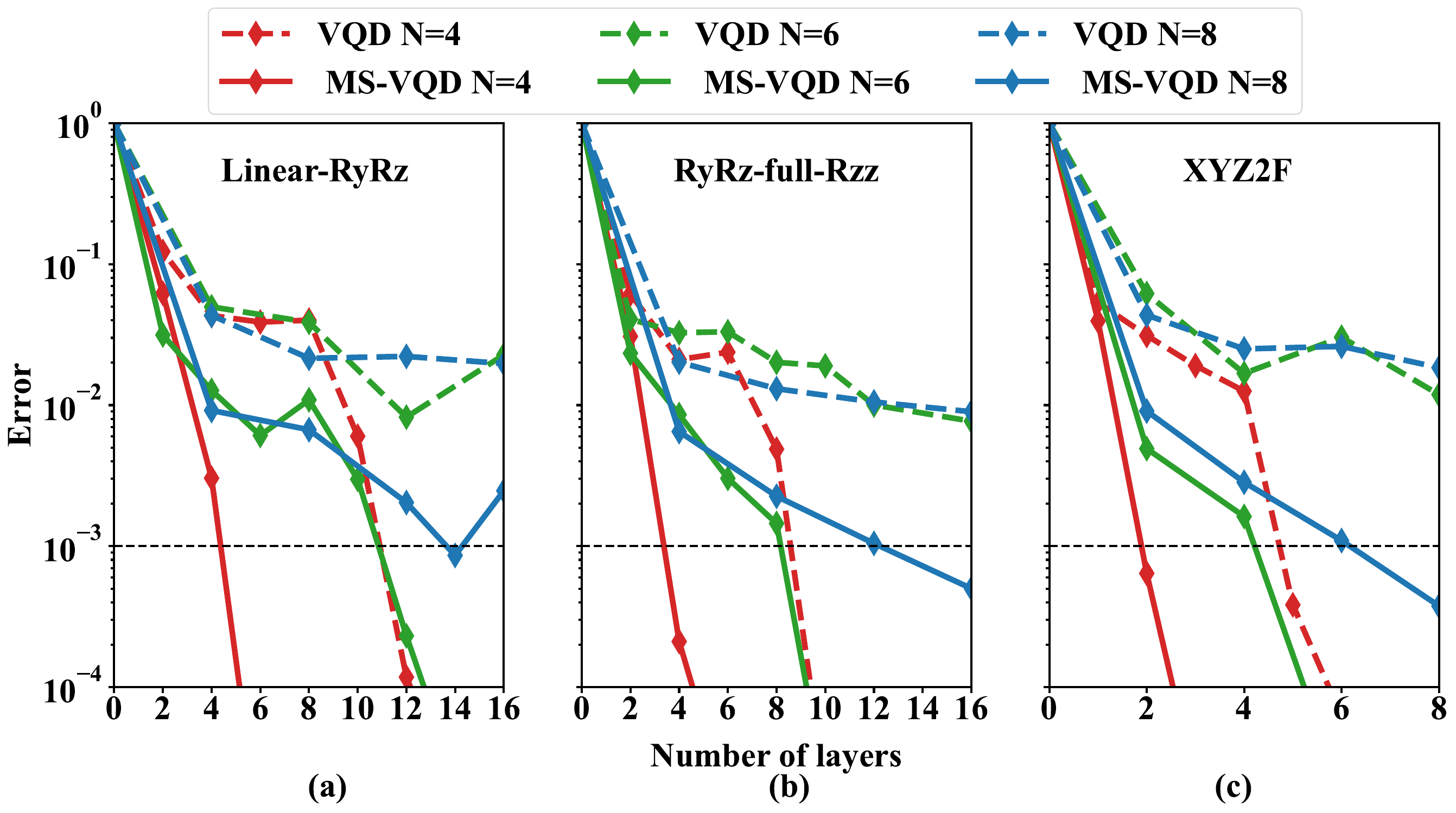}
    \caption{
    The population error for different layers of ansatz in
    VQD and MS-VQD. (a) Linear-RyRz ansatz, (b) RyRz-full-Rzz
    ansatz, and (c) XYZ2F ansatz. Different colors represent
    different system sizes.
    }
    \label{fig:error_vs_N}
\end{figure}

To quantitatively assess the performance of MS-VQD and VQD
across different system sizes, we defined the population error
as
\begin{gather}
    \varepsilon = \frac{\sum_{k=1}^{N_\textrm{steps}}
    |P_1 (t_k)
    - P_1^{\textrm{ref}}(t_k) |}{N_\textrm{steps}}.
    \quad t\in [0,10/\omega]
\end{gather}
Figure~\ref{fig:error_vs_N}(a-c) illustrate how the error
evolves with an increasing number of ansatz layers $L$ for the
three different ansatzes (The error at $L=0$ is arbitrarily set
to 1 for eye guidance). Regardless of the ansatz used,
MS-VQD exhibits faster convergence with $L$
compared to VQD. For example, in the case of an octamer using
the linear-RyRz ansatz, MS-VQD with $L=4$ achieves higher
accuracy than VQD with $L=16$. Similar trends are observed for
both the RyRz-full-Rzz and XYZ2F ansatzes, suggesting that the
advantage of MS-VQD is likely generalizable across other ansatz
choices.  When comparing the three ansatzes within MS-VQD, the
RyRz-full-Rzz ansatz performs slightly better than the
linear-RyRz ansatz in terms of PQC layers. For the octamer with
$L=16$, the error of the RyRz-full-Rzz ansatz is $5.0 \times
10^{-4}$, while that of the linear-RyRz ansatz is $2.0 \times
10^{-3}$. The XYZ2F ansatz performs much better,
achieving an error of $3.8 \times 10^{-4}$ with only $L=8$. This
trend is consistent with the findings from ground-state
calculations.~\cite{xiao2024physics}
We also used the wavefunction infidelity
$1-\textrm{Re}\langle\Psi^\textrm{ref}(t) |\Psi(t)\rangle$ as
the error metric, the findings are similar as shown in
Figure~S5.

\begin{figure}[htb]
    \centering
    \includegraphics[width=0.47\linewidth]{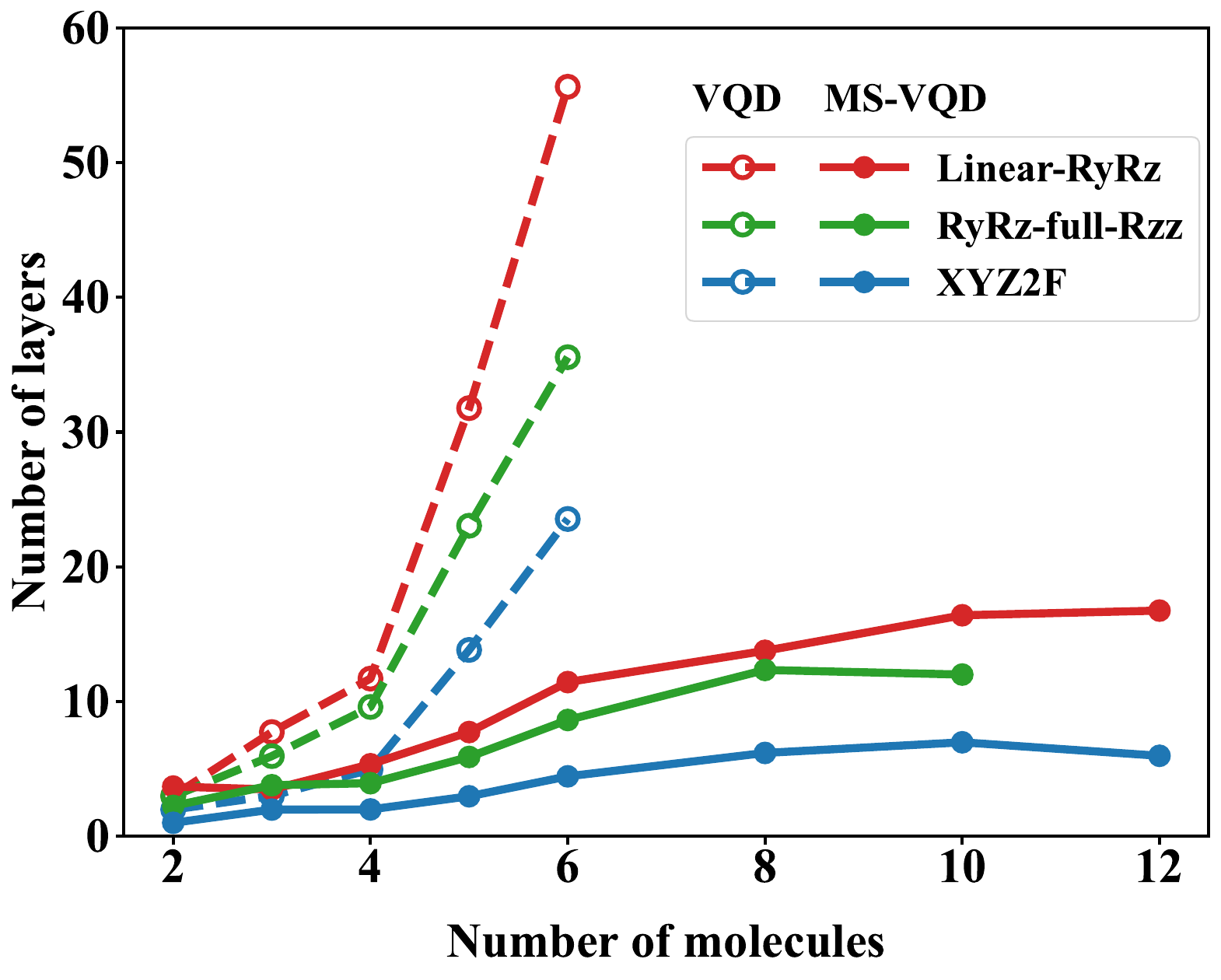}
    \caption{
    The required number of ansatz layers $L$ for different system sizes to achieve an error below $10^{-3}$. Solid lines represent MS-VQD results, while dashed lines represent VQD. Different colors denote different ansatzes.
    }
    \label{fig:L_vs_N}
\end{figure}

Figure~\ref{fig:L_vs_N} summarizes the number of
layers $L$ required to achieve an population error below $10^{-3}$ for
different system sizes. As system size increases, VQD struggles
to maintain accuracy, requiring a rapidly growing number of
layers. In contrast, MS-VQD scales much more favorably,
requiring far fewer layers to achieve the same accuracy. This
trend is consistent across all three ansatzes, further
highlighting the superiority of MS-VQD for nonadiabatic dynamics
involving a large number of electronic states.
Regarding the number of parameters in the ansatzes, MS-VQD
requires slightly fewer parameters than VQD (shown in Figure~S6), meaning
its PQC depth is approximately $1/N$ of that in VQD to achieve
the same accuracy.
This finding aligns with the schematic diagram in Figure~\ref{fig:pes}, illustrating that VQD requires $N$ ansatz blocks to capture the dynamics of $N$ electronic states accurately.
However, the number of parameters only
becomes smaller than the size of the full Hilbert space when the
system is larger than hexamer (Figure~S7), indicating that
quantum advantage may only emerge for studying large systems.

\begin{figure}[htb]
    \centering
    \includegraphics[width=0.5\linewidth]{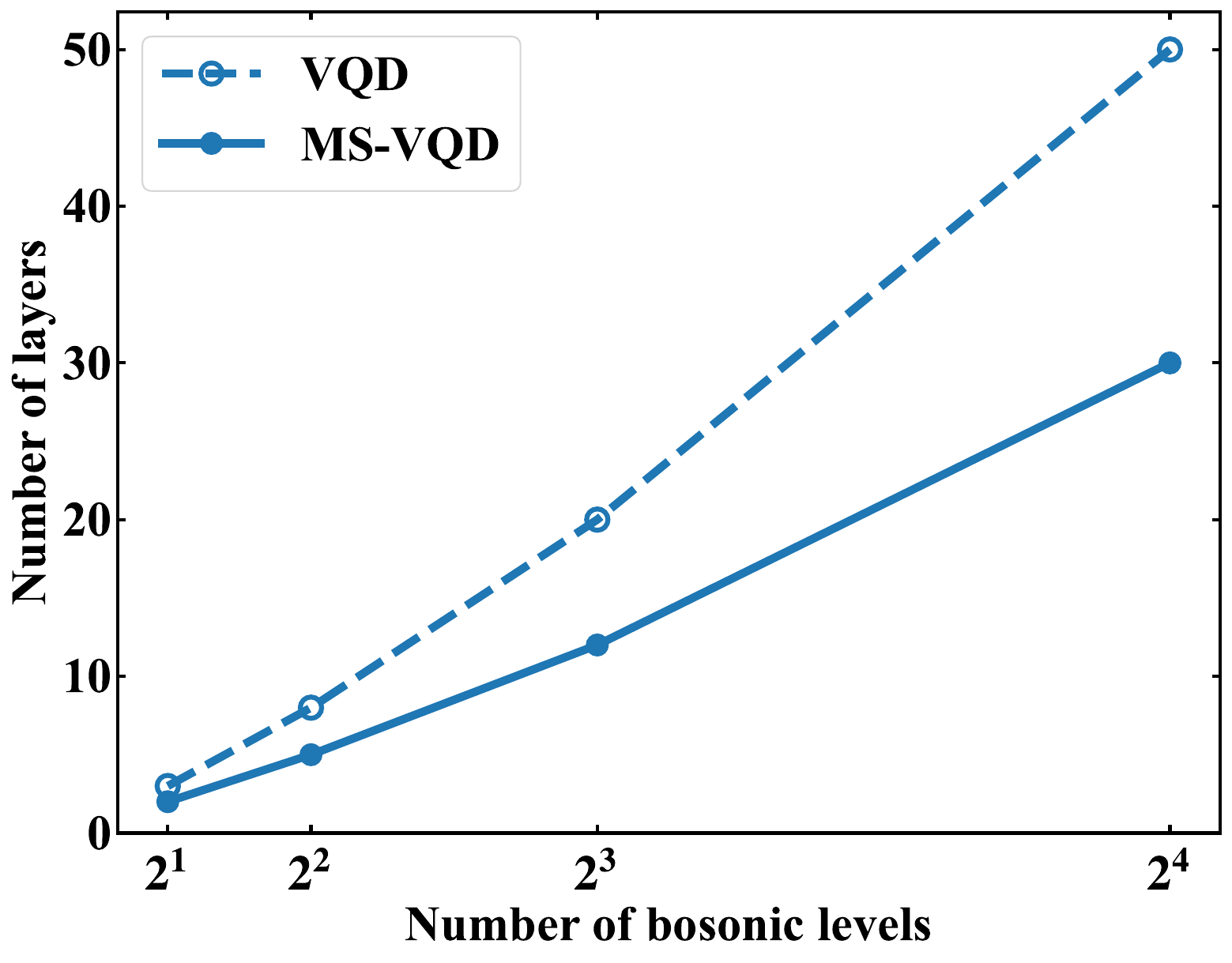}
    \caption{
    The required number of ansatz layers to achieve an error below $10^{-3}$ for different numbers of local bosonic
    levels $d$.
    }
    \label{fig:qn}
\end{figure}

Due to the exponentially large computational cost of classical
simulations of quantum computing, the above calculations for
large molecular aggregates are restricted to local bosonic
levels $d=2$. However, this truncation is not sufficient for
accurately representing real molecular aggregates. Therefore, we
evaluated the performance of VQD and MS-VQD with different $d$
on a molecular dimer. After mapping the Hamiltonian to qubits,
the total number of terms increases linearly with $d$ (Figure~S1(a)),
indicating that a larger $d$ will result in a more entangled state and thus deeper PQC. 
Since MS-VQD involves fewer Hamiltonian terms than VQD, it is
expected that MS-VQD requires a shallower PQC.
As shown in Figure~\ref{fig:qn}, as $d$ increases, the number of
ansatz layers required for MS-VQD to achieve an error below
$10^{-3}$ grows much more slowly compared to VQD. This suggests
that MS-VQD maintains its computational advantage even with
large local bosonic space.

Furthermore, we examined the robustness of MS-VQD across
different coupling regimes and varying numbers of vibrational
modes, as detailed in SI Section~6. Figure~S8 and S9
demonstrated that the advantage of MS-VQD over VQD remains
robust across different electronic and electron-vibration
coupling strengths, as well as for multiple numbers of
vibrational modes.

\begin{table}[htb]
    \caption{
    The circuit depth and the number of measurements required to
    calculate the matrices $\mathbf{A}$, $\mathbf{D}$,
    $\mathbf{B}$, and $\mathbf{H}$.
    }
    \label{tab:scaling}
    \centering
    \begin{tabular}{cccccc}  
        \toprule
        Algorithm              &      & $A_{lk}^{pp}$   & $D^{pp}
        _k$ & $B_l^{pq}$ & $H_{pq}$ \\
        \midrule
        \multirow{2}{*}{VQD}   &depth & $\mathcal O(L)$ & $\mathcal O(L)$ & $\mathcal O(L)$ & $\mathcal O(L)$   \\
                               &times & $\mathcal O(n_\theta^2)$
                               & $\mathcal O(n_\theta)$ &
                               $\mathcal O(fn_\theta) $ & $\mathcal O(f)$   \\            
        \multirow{2}{*}{MS-VQD}&depth & $\mathcal O(L/N)$ & $\mathcal O(L/N)$ & $\mathcal O(\beta L/N)$ & $\mathcal O(\beta L/N)$   \\
                               &times & $\mathcal O(n_\theta^2/N)
                               $ & $\mathcal O(n_\theta)$ &
                               $\mathcal O(fn_\theta/N)$ & $\mathcal O(f)$   \\
        \bottomrule
    \end{tabular}
\end{table}

Finally, we analyzed the computational cost of VQD and MS-VQD.
The scaling of circuit depth and the number of measurements with
respect to the number of ansatz layers $L$ and the number of parameters
$n_\theta$ is summarized in Table~\ref{tab:scaling}. 
$N$ is the number of electronic states.
For VQD, using either direct measurement or the
Hadamard test algorithm, the circuit depth required to compute
$A_{lk}$, $D_{k}$, $B_{l}$, and $E$ is $\mathcal{O}(L)$.
The corresponding
number of measurements required is $\mathcal{O}(n_\theta^2)$,
$\mathcal{O}(n_\theta)$, $\mathcal{O}(fn_\theta)$, and $\mathcal{O}(f)$,
respectively, where $f$ is the number of Hamiltonian terms after
mapping to qubits. 
In comparison, assuming that the number of parameters in MS-VQD
is the same as that in VQD to achieve the same accuracy, the ansatz layers required for MS-VQD 
is $1/N$ of that for
VQD. Hence, the circuit depth required to measure the electronic diagonal
elements $A_{lk}^{pp}$ and $D_k^{p}$ in MS-VQD is both
$\mathcal{O}(L/N)$. However, for the electronic non-diagonal
elements $B_l^{pq}$ and $H_{pq}$, all gates with parameters must
be controlled by an ancilla qubit in the Hadamard test,
introducing an overhead prefactor $\beta$. For example, a
controlled-Ry gate can be decomposed into two Ry gates plus two
CNOT gates, giving $\beta=4$. This results in a circuit depth of
$\mathcal{O}(\beta L/N)$. 
The number of measurements required to
obtain $A_{lk}^{pp}$ and $D_k^p$ is $\mathcal{O} ((n_\theta/N)^2 \times
N) = \mathcal{O}(n_\theta^2/N)$ and $\mathcal{O}(n_\theta/N \times N) =
\mathcal{O}(n_\theta)$, respectively. The number of
measurements required for $H_{pq}$ seems to be $\mathcal{O}
(fN^2)$, but since only the subset $\hat{T} + \hat{V}_{pq}$ of
the whole Hamiltonian $\hat H$ needs to be measured for each
$pq$ pair, the overall cost remains $\mathcal{O}(f)$. A similar
argument applies to $B_l^{pq}$, which requires $\mathcal{O}
(fn_\theta/N)$ measurements. 
Thus, while the equation of motion
for MS-VQD seems more complex than that of VQD, both the
circuit depth and the number of measurements of some matrix
elements benefit from a $1/N$ reduction. This advantage becomes
particularly useful as the number of electronic states
increases.

In this letter, we proposed a novel multi-set variational
quantum dynamics algorithm for simulating nonadiabatic quantum
dynamics and derived the equations of motion within the
framework of McLachlan's time-dependent variational principle
for the first time. This is the main contribution of our work.
The MS-VQD algorithm utilizes a set of parameterized quantum
circuits to represent the entire electronic-nuclear coupled
wavefunction, where each PQC corresponds to the wavepacket on a
single potential energy surface. This structure allows the
ansatz to better adapt to the landscape of different PESs,
making the PQCs more compact. Through simulations of the
Frenkel-Holstein electron-vibration coupling model, we found
that the ansatz depth required by MS-VQD is approximately
reduced to $1/N$ of that in VQD while maintaining the same
accuracy. This reduction in quantum resource
requirements makes MS-VQD more suitable for implementation on
current noisy quantum computers. Moreover, the advantage of
MS-VQD over VQD is robust across different ansatz choices, sizes
of the local bosonic space,
coupling regimes, and number of vibrational modes. In the future, MS-VQD
can also be extended to imaginary-time evolution~\cite{mcardle2019variational} for calculating
quantum statistical properties. Additionally, it can be combined
with grid basis to study nonadiabatic dynamics on real
anharmonic potential energy surfaces.~\cite{kassal2008polynomial, ollitrault2023quantum}
Despite its advantages, some challenges remain. Although MS-VQD
outperforms VQD in simulating nonadiabatic dynamics, achieving
highly accurate results still requires a large number of
parameters. For large molecular aggregates that may exhibit quantum advantage, the quantum
resources required by all three hardware-efficient ansatzes
explored in this study have exceeded the capabilities of current noisy
quantum computers. Given MS-VQD's strength in adapting to
wavepackets on each PES, exploring independent and
adaptive ansatz~\cite{grimsley2019adaptive,yao2021adaptive,
zhang2023low} for each PQC to replace the shared and fixed
ansatz may further compact the circuit and enhance efficiency.

\section*{Appendix: Derivation of MS-VQD} 
\label{app:appendix1}

Let any vector on the tangent space of $|\Psi(\vec{\theta}(t))
\rangle$ be denoted as $\varphi$. According to McLachlan’s
time-dependent variational principle, the optimal solution for
$\dot{\Psi}(\vec{\theta}(t))$ is the vector $\varphi$ that
minimizes the functional $\mathcal{F}$,
\begin{gather}
\mathcal F = \|\varphi+i\hat H \Psi(\vec{\theta}(t)) \|^2
\end{gather}
The minimal point satisfies
\begin{equation}
\delta \mathcal F = \langle \delta \varphi | \dot{\Psi}(\vec{\theta}(t)) + i \hat
    H\Psi(\vec{\theta}(t))\rangle + \textrm{h.c.} = 0
\label{eq:deltaF}
\end{equation}
For the multi-set wavefunction ansatz,
\begin{gather}
    \dot{\Psi}(\vec{\theta}(t))= \sum_q \dot{c}_q \phi_q \chi_q
    + \sum_{qk} c_q \phi_q \frac{\partial\chi_q}
    {\partial\theta^q_k}
    \dot{\theta}_k^q
\label{eq:deri1} \\
    \delta \varphi=\sum_{p} \delta a_{p}\phi_{p} \chi_p
    +\sum_{pl} c_{p}\phi_p
    \frac{\partial\chi_{p}}{\partial\theta^{p}_{l}} \delta b^{p}
    _{l} \label{eq:deri2}
\end{gather}
Note that while $\delta a_{p}$ can be complex, $\delta b^{p}_{l}
$ must be real, as required by the quantum computer.
Substituting eq~\eqref{eq:deri1} and \eqref{eq:deri2} into
eq~\eqref{eq:deltaF}, we obtain
\begin{gather}
    \left\langle\sum_p \delta a_p \phi_p \chi_p +\sum_{pl}
     c_p \phi_p \frac{\partial\chi_p} {\partial\theta^p_l}\delta
    b^p_l \middle | 
    \sum_q \dot{c}_q \phi_q \chi_q +\sum_{qk}
    c_{q} \phi_q \frac{\partial\chi_q}{\partial\theta^q_k}
    \dot{\theta}^q_k
    +i \hat H \sum_q c_q\phi_q\chi_q \right\rangle +
    \textrm{h.c.}=0
\end{gather}
The above formula can be separated into three parts: $\delta
a_p$, $\delta a_p^*$ (since $\delta a_p$ is complex, $\delta
a_p$ and $\delta a_p^*$ can be treated as independent
variations), and $\delta b_{l}^p$. Using the relation $\langle
\phi_p | \phi_q \rangle = \delta_{pq}$, these three parts are
given by:
\begin{enumerate}
\item $\delta a_p$ part:
\begin{gather}
    \dot c_p + c_p \sum_k D_k^{pp*} \dot \theta^p_k + i\sum_q
    c_q H_{pq}= 0
\label{eq:eq_c} 
\end{gather}

\item $\delta a_p^*$ part:
\begin{gather}
\dot c_p^* + c_p^* \sum_k D_k^{pp} \dot \theta^p_{k} - i\sum_q
    c^*_q H_{pq}^* = 0 \label{eq:eq_cstar}
\end{gather}
Comparing eq~\eqref{eq:eq_c} and eq~\eqref{eq:eq_cstar},
we see that if one expression holds, the other must also
be valid. Therefore, it is sufficient to consider only
one of them.

\item $\delta b^p_l$ part:
\begin{gather}
 c_p^*\left\langle\frac{\partial\chi_p}{\partial\theta^p_l}
    \middle|\dot c_p \chi_p\right\rangle + c_p^*
    \left\langle\frac{\partial\chi_p}{\partial\theta^p_l}
    \middle|\sum_k c_p\frac{\partial \chi_p}{\partial \theta^p_k}
    \dot \theta^p_k\right\rangle+
    c_p^*\left\langle\frac{\partial\chi_p}{\partial\theta^p_l}
    \middle|i \sum_q (\hat T + \hat V_{pq}) c_q \chi_q
    \right\rangle + \textrm{h.c.} = 0
\label{eq:eq_b}
\end{gather}
Substituting $\dot c_p$ from eq~\eqref{eq:eq_c} into eq~\eqref{eq:eq_b}, 
        we obtain eq~\eqref{eq:theta_deri} in the main text.

\end{enumerate}

\begin{acknowledgement}

    This work is supported by the Innovation Program for Quantum
    Science and Technology (Grant No. 2023ZD0300200), NSAF
    (Grant No. U2330201), the National Natural Science
    Foundation of China (Grant No. 22273005), and the
    Fundamental Research Funds for the Central Universities.    

\end{acknowledgement}

\begin{suppinfo}

    The circuits to measure matrix elements in MS-VQD;
    The number of Hamiltonian terms after mapping to qubits;
    The population dynamics with RyRz-full-Rzz ansatz and XYZ2F ansatz;
    Infidelity as a function of the number of ansatz layers;
    The number of parameters in MS-VQD and VQD;
    The performance of MS-VQD and VQD with different coupling strength and number of modes.

\end{suppinfo}

\bibliography{references}

\end{document}